\documentclass[11pt,twoside]{article}

\usepackage{asp2014}

\aspSuppressVolSlug
\resetcounters

\begin{document}

\title{Advanced Data Reduction for the MUSE Deep Fields}

\author{Simon~Conseil$^1$, Roland~Bacon$^1$, Laure~Piqueras$^1$, and Martin~Shepherd$^1$
\affil{$^1$Univ Lyon, Univ Lyon1, Ens de Lyon, CNRS, Centre de Recherche
Astrophysique de Lyon UMR5574, F-69230, Saint-Genis-Laval, France;
\email{simon.conseil@univ-lyon1.fr}}}

\paperauthor{Simon~Conseil}{simon.conseil@univ-lyon1.fr}{}{Univ Lyon, Univ Lyon1, Ens de Lyon, CNRS, Centre de Recherche Astrophysique de Lyon UMR5574}{}{Saint-Genis-Laval}{}{69230}{France}
\paperauthor{Roland~Bacon}{roland.bacon@univ-lyon1.fr}{}{Univ Lyon, Univ Lyon1, Ens de Lyon, CNRS, Centre de Recherche Astrophysique de Lyon UMR5574}{}{Saint-Genis-Laval}{}{69230}{France}
\paperauthor{Laure~Piqueras}{laure.piqueras@univ-lyon1.fr}{}{Univ Lyon, Univ Lyon1, Ens de Lyon, CNRS, Centre de Recherche Astrophysique de Lyon UMR5574}{}{Saint-Genis-Laval}{}{69230}{France}
\paperauthor{Martin~Shepherd}{martin.shepherd@univ-lyon1.fr}{}{Univ Lyon, Univ Lyon1, Ens de Lyon, CNRS, Centre de Recherche Astrophysique de Lyon UMR5574}{}{Saint-Genis-Laval}{}{69230}{France}

\begin{abstract}

The Multi Unit Spectroscopic Explorer (MUSE) is an integral-field spectrograph
operating in the visible wavelength range, and installed at the Very Large
Telescope (VLT). The official MUSE pipeline is available from ESO. However,
for the data reduction of the Deep Fields program (Bacon et al., in prep.), we
have built a more sophisticated reduction pipeline, with additional reduction
tasks, to extend the official pipeline and produce cubes with fewer
instrumental residuals.

\end{abstract}

\section{Introduction
\label{intro}}

MUSE is composed of 24 Integral-field spectrographs (IFU) operating in the
visible wavelength. The instrument has a field of view of $1' \times 1'$
sampled at 0.2 arcsec, an excellent image quality (limited by the 0.2 arcsec
sampling), a large simultaneous spectral range (4650--9300 \.{A}), a medium
spectral resolution ($R \simeq 3000$) and a very high throughput.

The data reduction for this instrument is the process which converts raw data
from the 24 CCDs into a combined datacube (with two spatial and one wavelength
axis) which is corrected for instrumental and atmospheric effects. Since the
instrument consists of many subunits (24 integral-field units, each slicing
the light into 48 parts, i.e. 1152 regions with a total of almost 90000
spectra per exposure), this task requires many steps and is computationally
expensive, in terms of processing speed, memory usage, and disk input/output.
This can be achieved with the MUSE standard pipeline
\citep{doi:10.1117/12.925114}, with is available from ESO
(\url{http://www.eso.org/sci/software/pipelines/muse/muse-pipe-recipes.html}).


For the data reduction of the \emph{Deep Fields} program (Bacon et al., in
prep.), we have built a more sophisticated reduction pipeline, with additional
data-reduction tasks, to extend the official one. We use this pipeline to
process the exposures we have for these two fields:
\begin{description}
    \item [Hubble Deep Field South (HDFS)] As one of the commissioning activities
        MUSE acquired single deep field in the HDFS. A $1' \times 1'$ field was
        observed to a 26.5h depth (53 $\times$ 1800 s).
        \citet{2015A&A...575A..75B} present a full description of the data,
        which is also available on the Muse Science
        website (\url{http://muse-vlt.eu/science/hdfs-v1-0/}).
    \item [Hubble Ultra Deep Field (UDF)] The MUSE-Deep GTO survey has observed
        a 9 field mosaic that covers the UDF. This $3' \times 3'$ field has been
        observed to a depth of $\sim$10h (in exposures of 1500 s each). In
        addition, there is also an extra-deep $1' \times 1'$ portion of the mosaic
        that reaches $\sim$31h.
        This data will be appear in Bacon at al. (in prep.), the
        redshifts in Brinchmann et al. (in prep.) and the full catalogue in Inami
        at al. (in prep.).
\end{description}


Section~\ref{processing} describes the data reduction pipeline that we
implement to process the data. Section~\ref{recipes} describes the additional
tasks that are used to reduce the data.





\begin{figure}[]
    \centering
    \includegraphics[width=\textwidth]{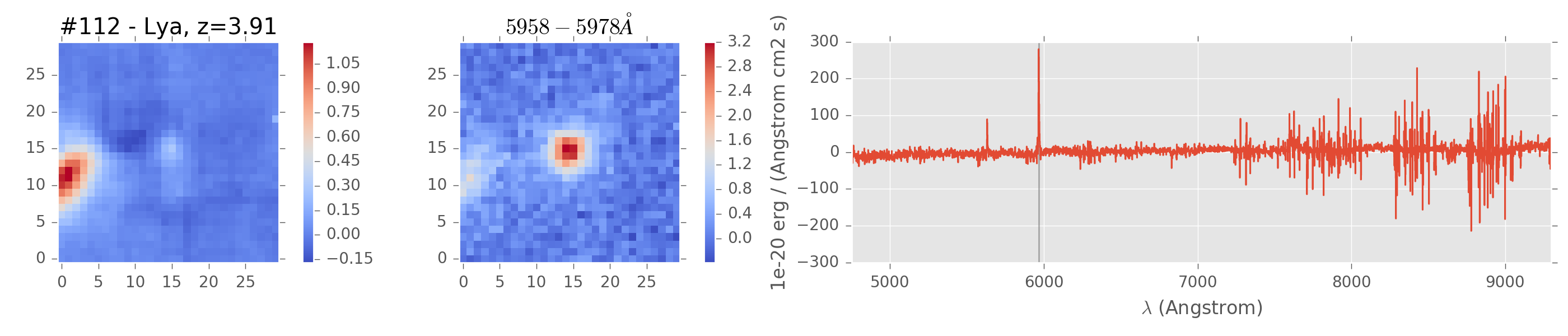}
    \caption{Example of a Ly$\alpha$ emitter detected in the HDFS data cube.}
\end{figure}

\section{Data Processing
\label{processing}}

When there are many exposures ($\sim$300 in our cases, $\sim$2.2Tb of raw
data), it becomes infeasible to manually keep track of everything. Instead it
is crucial to build automated and reproducible procedures for tasks like
associating the multiple calibration files needed for a given observation.
That's why we have built a custom data-reduction system with a database that
contains all of the information from the science and calibration exposures.
This lets us reliably identify files, based on criteria such as time offsets
or temperature differences.

Running all the reduction steps for all the exposures requires a lot of
computing time and produces a lot of files (several Tb per reduction).  To
keep track of the results and output logs of the different steps and versions,
we have built a processing pipeline based on several key components:%
\begin{itemize}
    \item \textbf{doit}: having a dozen of tasks to run for each exposure,
        one of the key points was to be able to track the state of the
        processing for each task.
        \emph{doit} (\url{http://pydoit.org}) is a task management and
        automation tool, open source and written in Python. \emph{doit} is
        a flexible tool which allows to define task dependencies, actions and
        targets. \emph{doit} remember the tasks execution, and can run
        multiple tasks in parallel.
    \item \textbf{SQlite}: a database used to store information about all the
        files (raw, calibration, outputs) and the runs.
    \item \textbf{Jupyter notebook}: used for the quality analysis.  It is an
        easy way to explore the data and plot the relevant information, while
        running directly onto the computing server and accessed remotely. Once
        the notebook is ready, we use it to generate HTML pages for each
        exposure.
\end{itemize}

\section{Additional reduction steps
\label{recipes}}

\begin{figure}[]
    \centering
    \includegraphics[width=\textwidth]{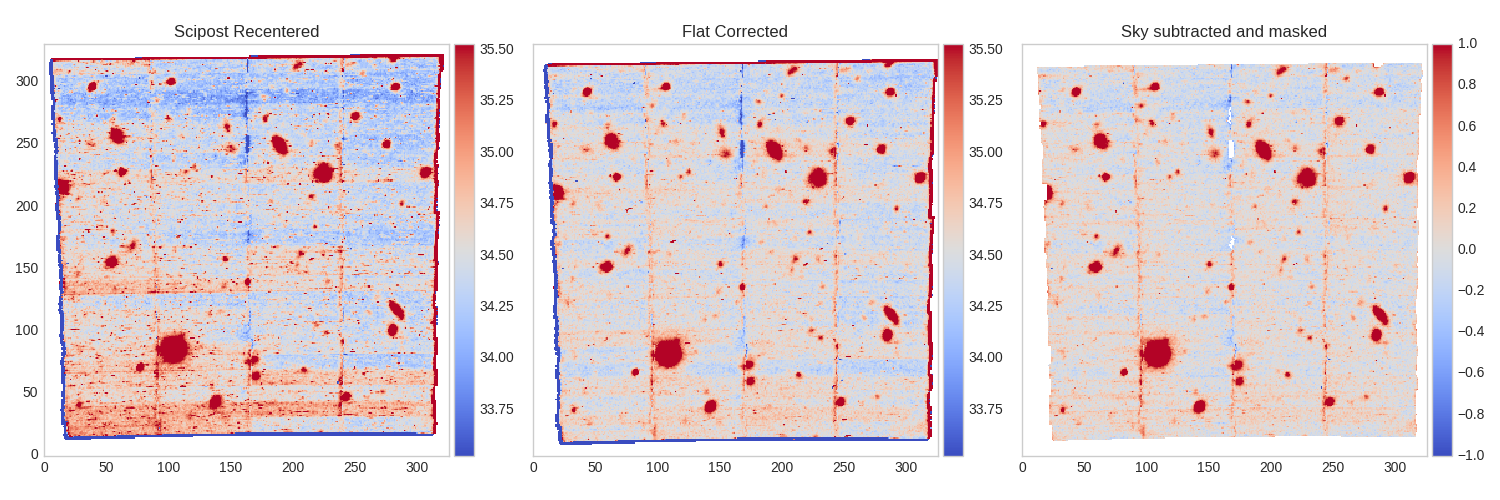}
    \caption{Reduction of an HDFS exposure: white-light image after running
        the pipeline (left), after recentering and correction of the slices
        level (middle), after the sky subtraction with ZAP and masking the
    edges (right).}\label{exp}
\end{figure}

The pipeline works well for the general case, but several things can be
improved with some additional reduction steps (Figure~\ref{exp}). Most of
these steps can be done with the recently released MPDAF Python package
\citep{P6.21_adassxxvi}.

\begin{figure}
    \centering
    \includegraphics[width=.78\textwidth]{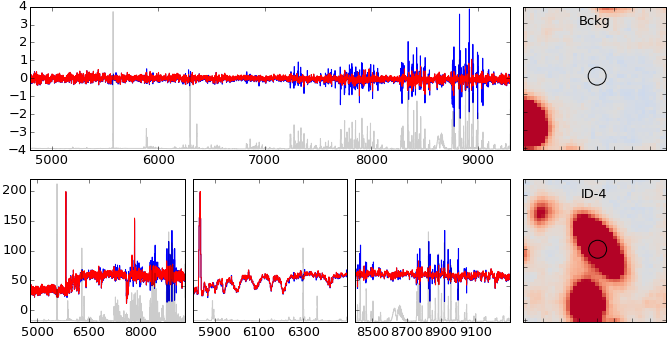}
    \caption{Spectra after ZAP (red) compared to the MUSE pipeline (blue)}\label{zap}
\end{figure}

\begin{figure}
    \centering
    \includegraphics[width=.78\textwidth]{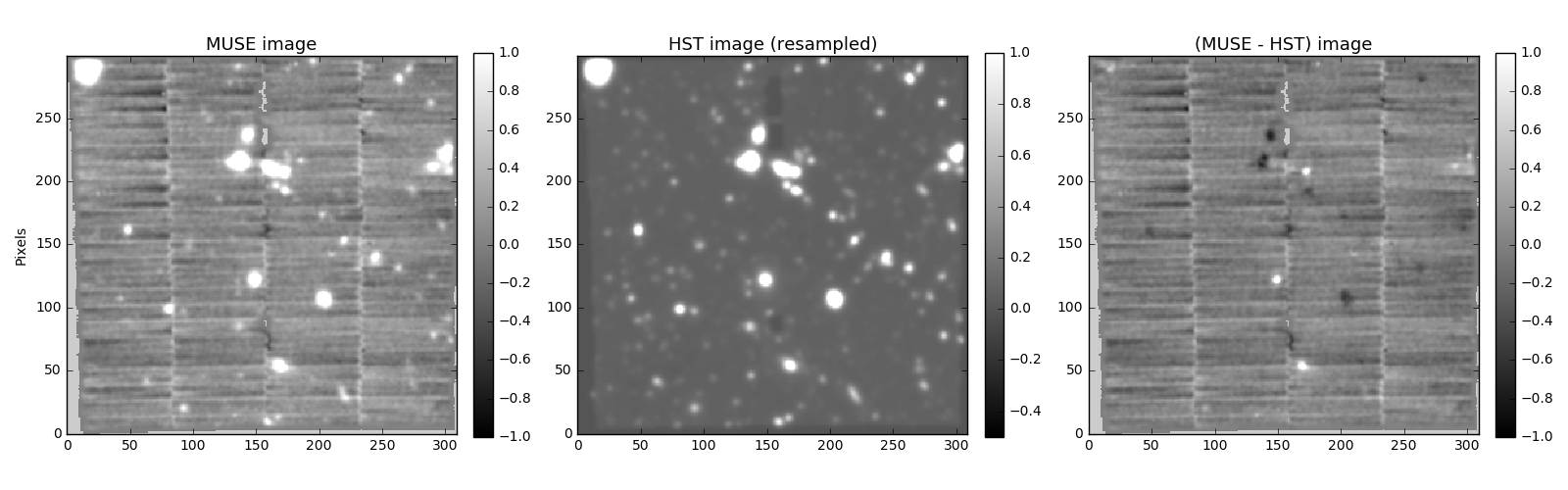}
    \caption{PSF estimation: fitted MUSE image, fitted HST image,
        and residual.}\label{psf}
\end{figure}

\begin{description}

    \item [Flat fielding] We developed an automatic flat-fielding procedure,
        which computes and apply a correction to the slices level, using the
        sky level as a reference (thus this needs to be done before the sky
        subtraction). Available in MPDAF\@:
        \small\begin{verbatim}
p = mpdaf.drs.PixTable('PIXTABLE.fits')  # open pixtable
mask = p.mask_column(maskfile=maskfile)  # mask sources
sky = p.sky_ref(pixmask=mask)            # compute the sky spectrum
cor = p.subtract_slice_median(sky, mask) # correct the slices level
p.write('PIXTABLE-COR.fits')             # save corrected pixtable
cor.write('AUTOCALIB.fits')              # save statistics
        \end{verbatim}\normalsize

    \item [Sky subtraction] Sky subtraction is performed with ZAP
        \citep{2016MNRAS.458.3210S}, a high precision sky subtraction tool,
        also released this year. The method uses Principal Component Analysis
        to isolate the residual features and remove them from the observed
        datacube (Figure~\ref{zap}). ZAP can also be run in addition to
        the sky subtraction of the MUSE pipeline.

    \item [Masking] Various masking steps are applied, to remove instrumental
        artifacts that cannot be corrected. This is done both on pixtables and
        datacubes.

    \item [Combination] Exposures combination applied on data cubes, which
        allow to run additional steps on the cubes before combining them. This
        is also part of MPDAF.  Several combination algorithm are available
        (mean, median, sigma clipping, with the \texttt{CubeList} class), and
        it can be used to create a mosaic (\texttt{CubeMosaic}).

    \item [PSF estimation] We developed a way to estimate the PSF parameters,
        calibration factors and pointing offsets, using HST as a reference:
        for each HST band, the image is resampled to the MUSE resolution, and
        fitted to a MUSE image computed on the same band (Figure~\ref{psf}).

\end{description}


\section{Conclusion
\label{conclusion}}

The additional data reduction tasks presented here allow us to produce final
(combined) cubes with reduced sky residuals, smaller background variations,
and fewer instrumental effects, which is useful for the next steps: sources
detection, redshifts estimation, and the scientific exploitation. Most of the
steps are available in the recently released MPDAF Python package
\citep{P6.21_adassxxvi}.

\acknowledgements RB acknowledges support from the ERC advanced grant MUSICOS.

\bibliography{P1-8}

\begin{thebibliography}{}
\expandafter\ifx\csname natexlab\endcsname\relax\def\natexlab#1{#1}\fi
\expandafter\ifx\csname url\endcsname\relax
  \def\url#1{\texttt{#1}}\fi
\expandafter\ifx\csname urlprefix\endcsname\relax\def\urlprefix{URL }\fi
\providecommand{\eprint}[2][]{\url{#2}}

\bibitem[{{Bacon} et~al.(2015){Bacon},  et~al.}]{2015A&A...575A..75B}
{Bacon}, R., , et~al. 2015, \aap, 575, A75. \eprint{1411.7667}

\bibitem[{{Piqueras} et~al.(2017){Piqueras}, {Conseil}, {Shepherd}, \&
  {Bacon}}]{P6.21_adassxxvi}
{Piqueras}, L., {Conseil}, S., {Shepherd}, M., \& {Bacon}, R. 2017, in ADASS
  XXVI, edited by TBD (San Francisco: ASP), vol. TBD of ASP Conf. Ser., TBD

\bibitem[{{Soto} et~al.(2016){Soto}, {Lilly}, {Bacon}, {Richard}, \&
  {Conseil}}]{2016MNRAS.458.3210S}
{Soto}, K.~T., {Lilly}, S.~J., {Bacon}, R., {Richard}, J., \& {Conseil}, S.
  2016, \mnras, 458, 3210. \eprint{1602.08037}

\bibitem[{{Weilbacher} et~al.(2012){Weilbacher}, {Streicher}, {Urrutia},
  {Jarno}, {P{\'e}contal-Rousset}, {Bacon}, \&
  {B{\"o}hm}}]{doi:10.1117/12.925114}
{Weilbacher}, P.~M., {Streicher}, O., {Urrutia}, T., {Jarno}, A.,
  {P{\'e}contal-Rousset}, A., {Bacon}, R., \& {B{\"o}hm}, P. 2012.
  \urlprefix\url{http://dx.doi.org/10.1117/12.925114}

\end{thebibliography}

\end{document}